\documentclass[conference]{IEEEtran}
\IEEEoverridecommandlockouts
\usepackage{cite}
\usepackage{amsmath,amssymb,amsfonts}
\usepackage{algorithmic}
\usepackage{graphicx}
\usepackage{textcomp}
\usepackage{xcolor}
\usepackage{soul}
\soulregister\cite7
\soulregister\ref7
\soulregister\pageref7
\soulregister\acrshort7
\soulregister\emph7
\usepackage{subfig}
\graphicspath{{img/}{img/_plots_marc/2023-8-24_11-40-49__2023-9-28_23-02-20}}
\DeclareGraphicsExtensions{.pdf,.jpeg,.png}
\def\BibTeX{{\rm B\kern-.05em{\sc i\kern-.025em b}\kern-.08em
    T\kern-.1667em\lower.7ex\hbox{E}\kern-.125emX}}
\begin{document}

\title{On the Impact of Inter-node Latency on Application Performance in Edge Networking\\
}

\author{\IEEEauthorblockN{Marc~Michalke,~Francisco~Carpio~and~Admela~Jukan}
    \IEEEauthorblockA{\\ \textit{Institute of Computer and Network Engineering,}
        \textit{Technische Universit{\"a}t Braunschweig}, Germany \\
        E-mail: \{m.michalke, f.carpio, a.jukan\}@tu-bs.de}
}


\maketitle

\begin{abstract}
When clustering devices at the edge, inter-node latency poses a significant challenge that directly
impacts the application performance. In this paper, we experimentally examine the impact that
inter-node latency has on application performance by measuring the throughput of an distributed
serverless application in a real world testbed. 
We deploy Knative over a Kubernetes cluster of nodes and emulate networking delay between them to
compare the performance of applications when deployed over a single-site versus multiple
distributed computing sites. 
The results show that multi-site edge networks achieve half the throughput
compared to a deployment hosted at a single site under low processing times conditions, whereas the
throughput performance significantly improves otherwise.
\end{abstract}


\section{Introduction}

Computing services today are increasingly hosted over geographically distributed sites in order to
reduce latency toward the client devices, which has lead to the wide deployment of smaller
multi-site clouds also referred to as distributed edge computing infrastructure. This new multi-site
topology of edge networking is different from traditional cloud computing.  In terms of
delay, single site clouds offer a negligible inter-node latency, while the client has to bridge a
significant geographical distance towards a distant cloud data center, leading to higher delays on
the corresponding connection link. On the other hand, edge computing offers comparably lower access
delays, but exhibits latency between nodes in the multi-site setting. The impact of
inter-node delay on application performance in a multi-site setting is still an open issue.



The inter-node latency in edge networking has been addressed by previous work. Works
\cite{Mittal_2021},\cite{Bracke_2023} propose a framework to addressing these issues, through the use
of new scheduling algorithms that co-locate multiple functions onto the same machine if they are
known to communicate with each other, thus reducing fragmentation and allowing for fairness \cite{Mittal_2021},\cite{Bracke_2023}. Other related work tackled the general
problem of latency-aware placement of workloads in \cite{Santos_2020} and \cite{Tajiri_2022}, while
the more concrete problem of latency consideration between the hosting edge nodes in a Kubernetes
environment was studied in\cite{Marchese_2022}, \cite{Wojciechowski_2021} and
\cite{Santos_2019}. These papers proposed to implementing Kubernetes scheduler plugins with
consideration for node-to-node latency.  Further experimental studies are still needed to show the
effects of link metrics like inter-node latency on the overall application throughput, and thus
practical application scaling. Experimental frameworks that were proposed to allow for metric
manipulation, without the need for specialized hardware like radio antennas, follow a softwarization
approach, such as in \cite{Fantom_2023} but are often limited to special areas like integration
testing in this example. Due to the labor-intensive process of creating such testbeds for
performance testing, the real-world implications of inter-node latency for distributed application
performance still have to be addressed which will be addressed by this work.



\begin{figure*}[!ht]
  \centering
  \subfloat[Benchfaster testbed]{\includegraphics[width=0.45\textwidth]{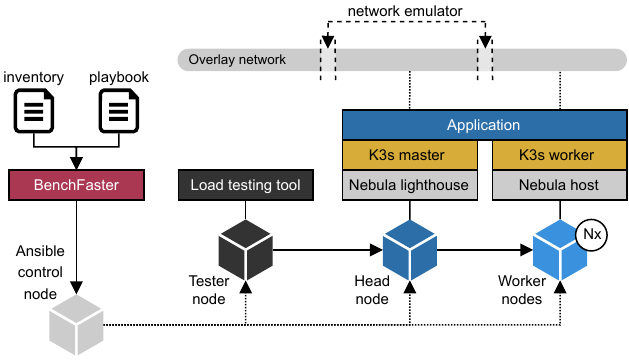}
      \label{fig:testbed}}
  \hfil \hspace{0.4cm}
  \subfloat[Single-site]{\includegraphics[width=0.225\textwidth]{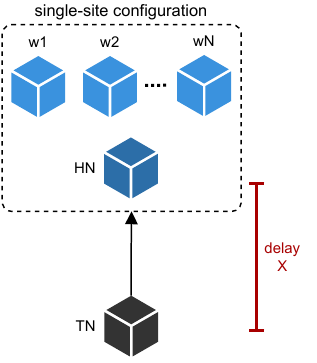}
  \label{fig:testbed_cs}}
\hfil
\subfloat[Multi-site]{\includegraphics[width=0.255\textwidth]{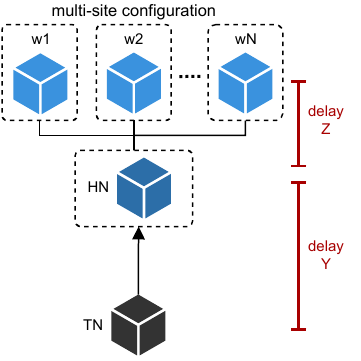}
\label{fig:testbed_ds}}
  \caption{\emph{Benchfaster} testbed and site configurations.}
\end{figure*}

In this paper, we contributed to the existing body of work with an experimental study of the impact that inter-node latency in a typical multi-site edge computing network environment. in a real-world testbed, we
measure the performance of a serverless application that scales across multiple computing sites by
measuring the achievable throughput.  We choose a serverless application example since it represents
a simple, distributed workload that can easily be implemented. We deploy our
open-source benchmarking and load testing testbed which was developed to allow for automated configuration of
various network metrics to impact deployed applications. The proposed testbed is based on Ansible
and is fully reproducible and automated,which is its salient feature. We measure the impact of latency on the overall throughput
of the application by adding select emulated delay values between nodes in order to compare their
performance when deployed over a single or multiple distributed sites. The results show how scaling
an application across multiple sites leads to approximately half the throughput of a comparable
single site as long as the processing times are shorter, which depends on the number of concurrent 
user requests. The measurements point to the need for an enhanced scheduling mechanism in Kubernetes and therefore also Knative, which should be latency-aware.


\section{The Open-Source Testbed: \emph{Benchfaster}} 

We introduce \emph{Benchfaster}\cite{benchfaster}, a performance evaluation framework used for the
realization of this work.  \emph{Benchfaster} is an evolution of our previous testbed described in
\cite{benchfaas}, with extended modularity to support generic containerized applications including
different serverless frameworks while also allowing for the use of different load testing tools. It
allows for an easy-to-use definition of different network topologies by adding delay between nodes
in the cluster and benchmarking containerized applications in an automated manner.

Fig. \ref{fig:testbed} shows \emph{Benchfaster} conceptually. It relies on Ansible, an automation
software that allows for definition of infrastructure as code by using sets of instructions, so
called playbooks, which can be structured hierarchically. Our framework consists of two of such
playbooks; a general one defining the creation of a cluster and a specific part added by the user,
defining the concrete applications to be deployed on top, with parameters for the infrastructure
creation and the performance tests to be executed. The machines to be used to form that
infrastructure are defined in an inventory file. All these parts, as well as Ansible itself, only
have to be deployed onto the control node which then provisions all other devices.

\subsection{Infrastructure and requirements}

\emph{Benchfaster} supports both, AMD64 and ARM64 architectures, as well as the deployment of nodes
as Virtual Machines (VMs). Each appliance, either physical or virtual, can be either (a) Ansible
control node, (b) Tester node,  (c) Headnode or (d) Worker nodes. The control node instructs all
machines through SSH on what and how to deploy the framework components.  The tester node conducts
the measurements once the setup process is completed and serves the results to the control node. The
headnode creates the cluster and runs the control plane for it while potentially also hosting
application containers. The worker nodes are managed by the headnode and deploy only application
containers as well as the components necessary to enable the management through the headnode. Each
node requires a Linux distribution, specifically Ubuntu Server 22.04 or Arch Linux. In the control
node, Ansible is required to be installed while the other nodes need to have passwordless sudo
permissions and key-based SSH access enabled.

\subsection{Overlay networking}

The testbed deploys \emph{Nebula} for overlay networking, which allows for communication between
the different nodes without having to take care of router or firewall configurations.
Specifically, the headnode hosts the \emph{Nebula} lighthouse which is then connected to by all the
worker nodes and therefore has to be accessible by not just the tester node but also all workers.

\begin{figure*}[!ht]
  \centering
  \includegraphics[width=0.9\textwidth]{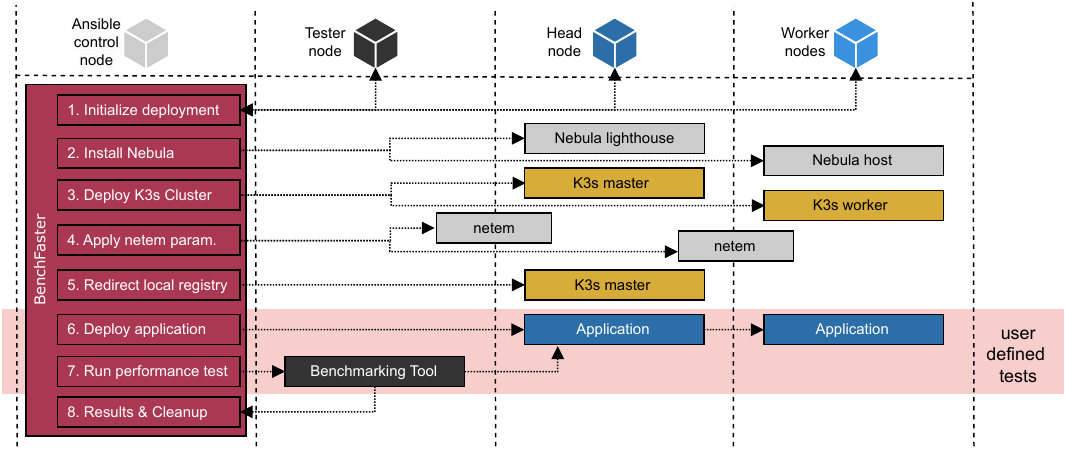}
  \caption{\emph{Benchfaster} deployment diagram}
  \label{fig:benchfaster_diagram}
\end{figure*}
\subsection{Container orchestration}

As container orchestrator we choose \emph{k3s}, a lightweight Kubernetes distribution designed for
resource-limited devices, which makes it compatible with most Kubernetes applications.  The
\emph{k3s} cluster is initiated and managed by the headnode, which also hosts the Kubernetes control
plane and then joined by all worker nodes through the overlay network. When the headnode later
receives a request to an application service, it will either forward the request further to a worker
that has a matching replica already deployed, or serve the request itself if there is an instance
present locally.

\subsection{Application}
The framework can deploy any containerized applications compatible with Kubernetes and scale them
according to Kubernetes' configuration options, for example through pod affinity and Horizontal Pod
Autoscaler (HPA) rules. Up to now, Knative, OpenFaas, Eclipse Mosquitto and RabbitMQ are tools that
have already been tested on \emph{Benchfaster}.

\subsection{Configuration of network metrics and parameters}\label{sec:qos_manipulation} To run the measurements,
\emph{Benchfaster} can modify the interfaces of the overlay network to configure link metrics like
packet loss, delay, packet corruption or jitter. This enables the user to emulate different network
environments by replicating the specific attributes of connections like geographical distance or
lossy wireless protocols. The framework can therefore emulate a cloud-like single site configuration
as depicted in Fig. \ref{fig:testbed_cs}, with delay $X$ only between the tester node and the
headnode and no extra delay between the nodes of the cluster which are assumed to be connected by
high-speed connections within the same data center. Emulation of multiple geographically distributed
sites is also possible, by using the configuration depicted in Fig. \ref{fig:testbed_ds}, where
delay is added not only between tester and the head node (delay $Y$), but also in-between the nodes
of the cluster (delay $Z$).  For the network metric manipulation, the framework relies on the
\emph{netem} package, part of all major Linux distributions, allowing for this kind of modification
of physical and virtual interfaces.

\subsection{Load testing tools} 

\emph{Benchfaster} supports \emph{JMeter}, its prospective successors \emph{K6} as well as
\emph{hey}, an open source program written in Golang, as load testing tools. Which load testing tool
to choose for the individual performance tests depends on the desired test metrics, the deployed
applications as well as the testbed engineer's- and developer's preferences.

 \subsection{Deployment workflow} 

Fig. \ref{fig:benchfaster_diagram} shows the deployment workflow. It starts from the Ansible control
node that \emph{Benchfaster} repository is cloned and the inventory file is added to. When
running one of the provided Ansible playbooks, \emph{Benchfaster} first initializes the different
parameters (see Step 1) such as the results folder with current timestamps, prepares Nebula
certificates, and cleans up previous deployments from all machines, in case they are still running.
In Step 2, the control node installs the Nebula lighthouse onto the headnode and the Nebula clients
onto the worker nodes which allows for reachability between all the machines in the cluster,
regardless of whether they are in the same network or not as long as the Nebula lighthouse is
publicly reachable. In Step 3, the K3s cluster is deployed by first installing K3s on the headnode
and then joining the worker nodes to the cluster as clients.  Step 4 consists of applying the link
metrics specified in the playbook to the corresponding network interfaces of the machines.  Step 5
whitelists a local container registry in the K3s configuration and replaces pointers to public
registries like Docker Hub with it in the deployment files. While this step is optional, it is
highly recommended, since it avoids limitations on the download rates imposed by public registries,
which may affect the total duration of the tests.  In Step 6, the application configured by the user
is deployed on top of the Kubernetes cluster and potential additional configuration files are
applied.  Step 7 is also fully user defined and can consist of additional control loops if there are
multiple dimensions for test parameters or multiple different tests to run against different
applications.  Once completed, the results from the tests aer generated by the load testing tool and retrieved in Step 8. The cleanup process resets all link metrics, uninstalls K3s and removes
Nebula.

\subsection{Design paradigms}

Since the whole framework is designed around non-interactive principles,
multiple repetitions to ensure reproducibility of the results can be performed
through Bash loops from the CLI. To eliminate influence between runs, the
deletion process removes the overlay network binaries from the headnode and all
workers, deletes any applied QoS settings and uninstalls the container
orchestration software, which also removes all components building on top of it,
including all deployed applications and their traces. The deployment and testing
procedure is then be repeated in a clean environment with the next set of
parameters if applicable. This procedure ensures that individual test results
can neither be impacted by multiple runs of the same test, nor by configuration
steps that are not part of the playbooks.
\begin{figure*}[!ht]
  \centering
  \includegraphics[width=0.9\textwidth]{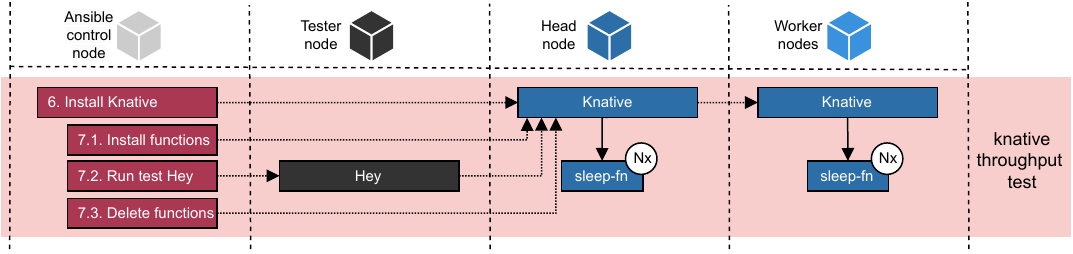}
  \caption{Specific test workflow}
  \label{fig:specific_diagram}
\end{figure*}

\section{Testbed configuration and setup}
In this section, we describe the testbed hardware, network and applications deployed, as well as the design of the test workflow for the measurements obtained.

\subsection{Hardware}
The testbed consists on four desktop computers. Three machines are used for the
cluster consisting of one headnode and two workers nodes, each of them being
equipped with with 4 cores and 16 GB of RAM. The fourth machine, the tester
node, consists of an Intel i7 6700 with 32 GB of RAM. All devices run Ubuntu
server 22.04.1 and are connected to a single Gigabit switch. The Ansible control
node has also access to all machines through the campus area network.

\subsection{Network}

We conduct the measurements across two different network scenarios, a cluster
across a single site and across multiple sites, representing a geographically
centralized and distributed environment, respectively. We then increase the
introduced delay per topology and conduct tests in each environment. Based on
the delay distribution explained in \ref{sec:qos_manipulation}, we configure the
delays to be $Z=Y=X/2$, where we evaluate different values of $X$ from 12.5ms to
800ms. In this way, the emulated total delay added between the end user (tester
node) and the application is comparable in both network scenarios.

\subsection{Serverless function and autoscaling}
We conduct measurements designed to quantify the serverless function throughput without involving
processing. In this way, we avoid that the choice of hardware impact the measurements.  To this end,
we first install the serverless platform \emph{Knative}, which offers support for the HPA, as well
as an option created by the Knative team, the Knative Pod Autoscaler (KPA). The latter adds
additional scaling and scheduling capabilities specifically designed for scaling serverless
functions.  On top of that, the test function is deployed which consists of a sleep process that is
being run for a duration of time which is specified in the payload of the request sent by the
client. This approach of having the program simply wait for an absolute amount of time instead of
performing processing operations ensures predictable durations regardless of the underlying hardware
capabilities. For a better understanding of the impact of the hardware, one can refer to the related
work in \cite{benchfaas}. The function can be deployed on top of \emph{Knative} to serve requests
and signals completion by returning a successful response. We published the source code for the test
function in a separate Github repository.\footnote{https://github.com/fcarp10/serverless-functions}

Scaling decisions for \emph{Knative} are made by KPA in the headnode. The
scaling for the \emph{sleep} function is not explicitly defined. It therefore
uses the default settings of KPA which are based on the \emph{concurrency} mode
with scale-to-zero (soft-limit: 100, hard-limit: 0, target utilization: 70,
stable window: 60, panic window: 10).  With these parameters, the function is
scaled up once Knative detects that the function utilizes more than 70\% of its
assigned resources within a window of 60 seconds, unless the autoscaler is in
"panic mode". This mode is entered whenever the current utilization reaches
200\%, which results in the window for measuring the average utilization being
reduced to 6 seconds. The soft limit defines a maximum of 100 concurrent
replicas as ceiling for the scaling while the scale-to-zero option defines a
minimum of zero replicas. The soft limit however can be exceeded if there are
sudden, extreme bursts of requests, which we do not consider here. 

\begin{figure*}[hbt!]
  \centering
  \subfloat[1 user - single-site]{\includegraphics[width=0.49\textwidth]{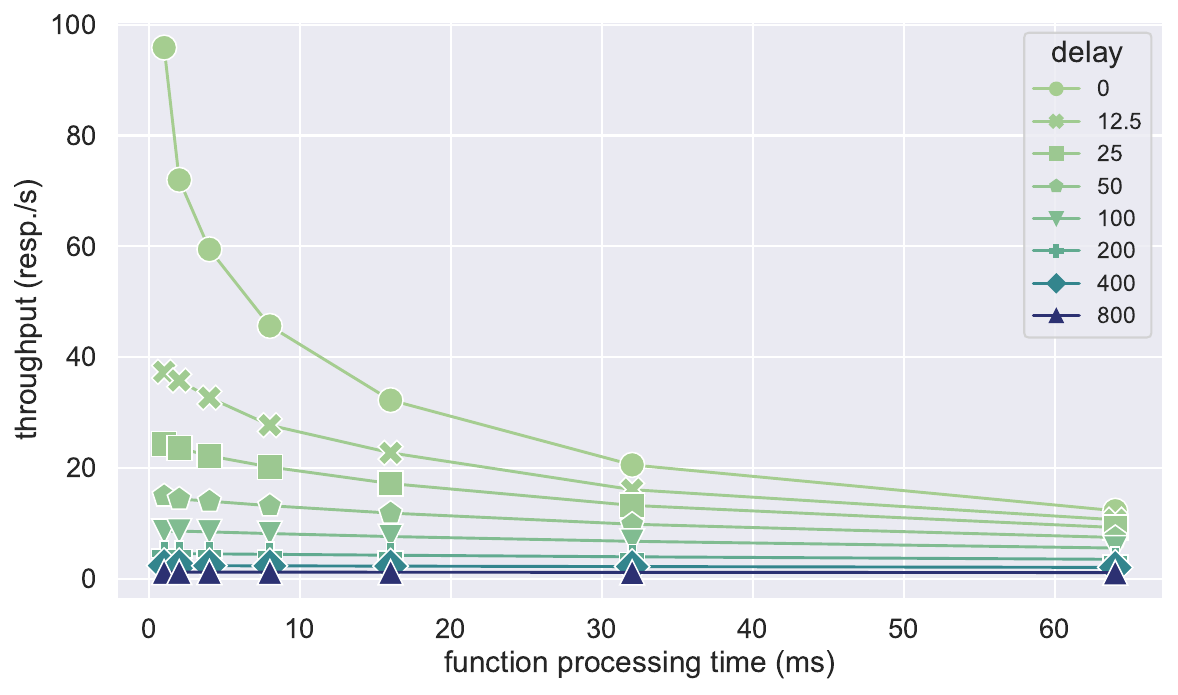}
      \label{fig:scalability-sleep-fn_CS_T-1}}
  \hfil
  \subfloat[1 user - multi-site]{\includegraphics[width=0.49\textwidth]{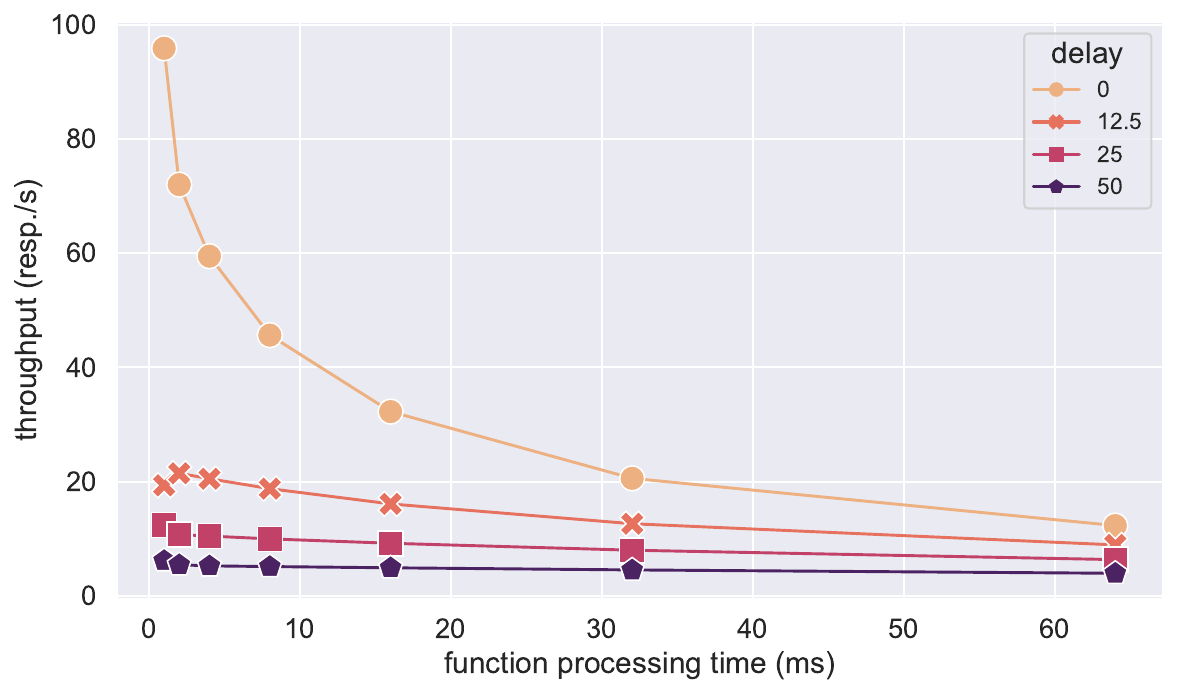}
      \label{fig:scalability-sleep-fn_DS_T-1}}
  \hfil
  \subfloat[50 users -  single-site]{\includegraphics[width=0.49\textwidth]{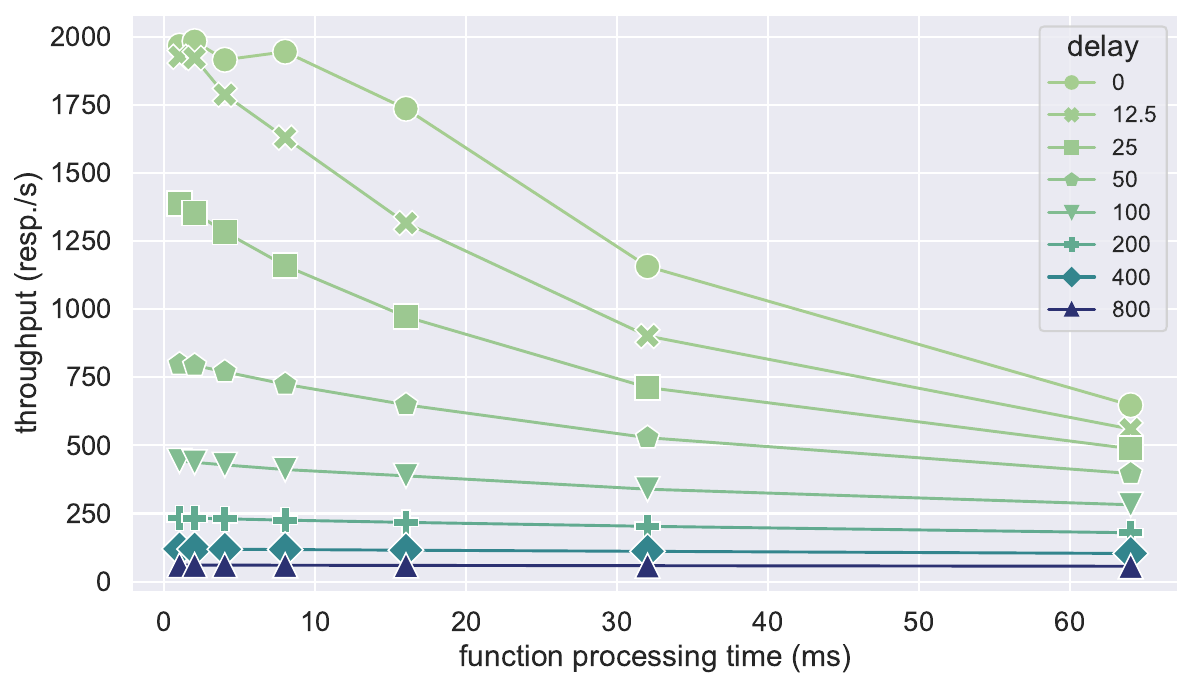}
  \label{fig:scalability-sleep-fn_CS_T-50}}
  \hfil
  \subfloat[50 users - multi-site]{\includegraphics[width=0.49\textwidth]{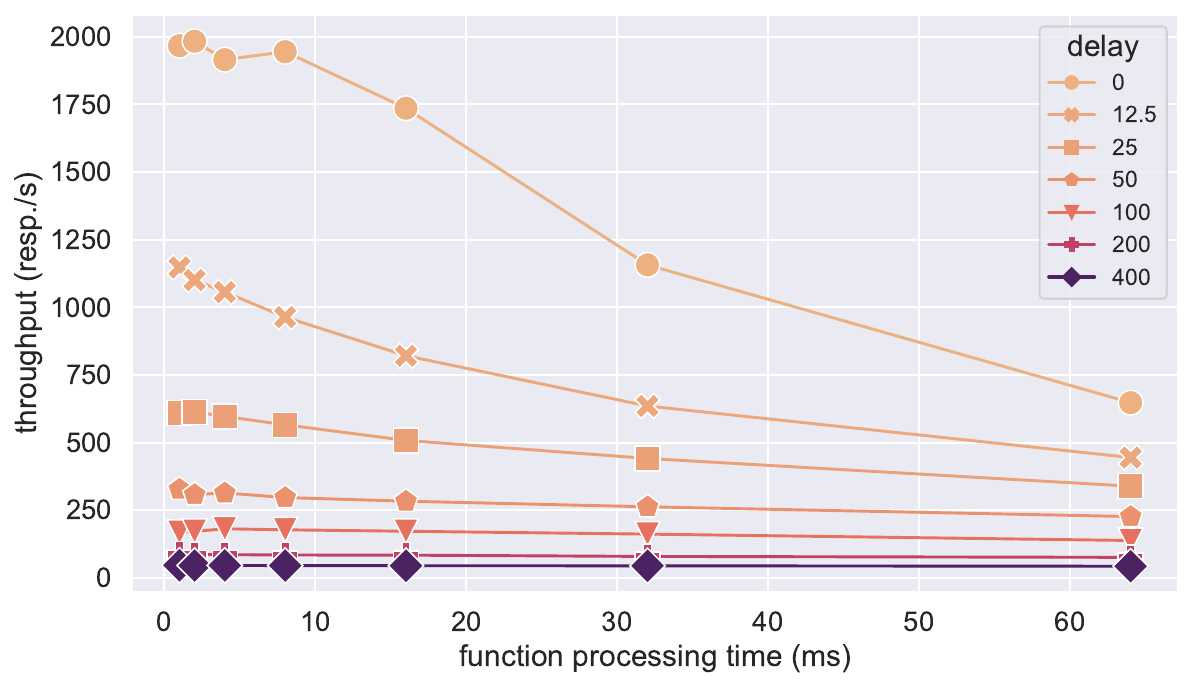}
  \label{fig:scalability-sleep-fn_DS_T-50}}
  \hfil
  \subfloat[500 users -  single-site]{\includegraphics[width=0.49\textwidth]{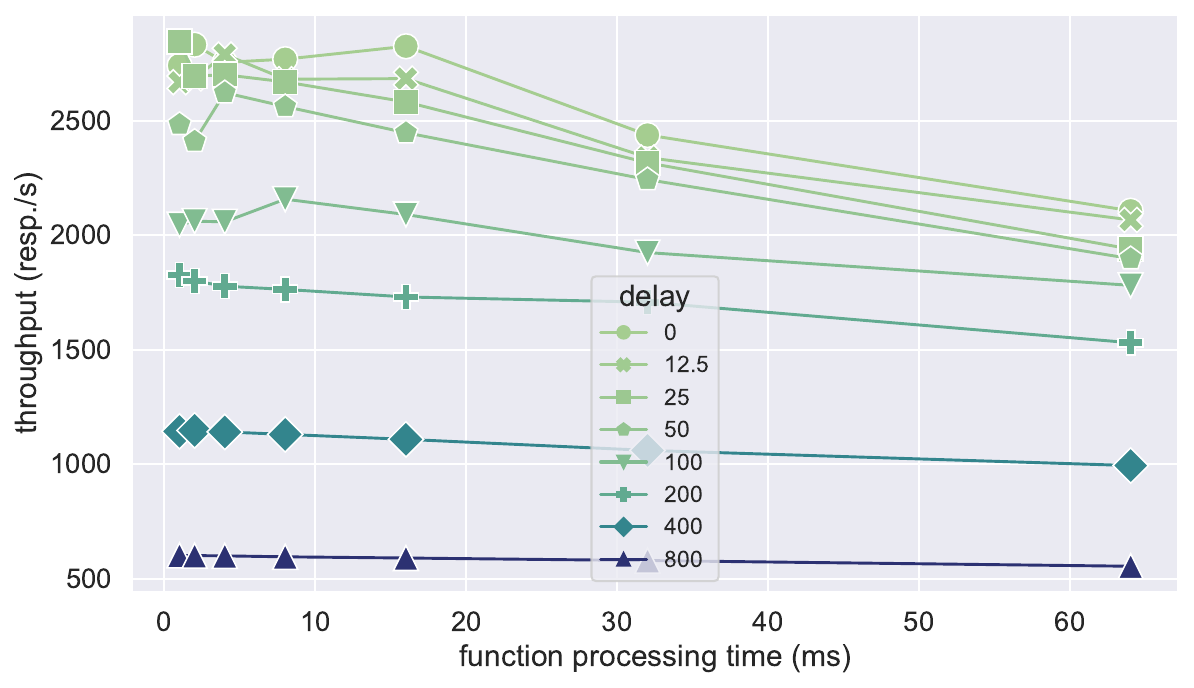}
\label{fig:scalability-sleep-fn_CS_T-500}}
  \hfil
\subfloat[500 users - multi-site]{\includegraphics[width=0.49\textwidth]{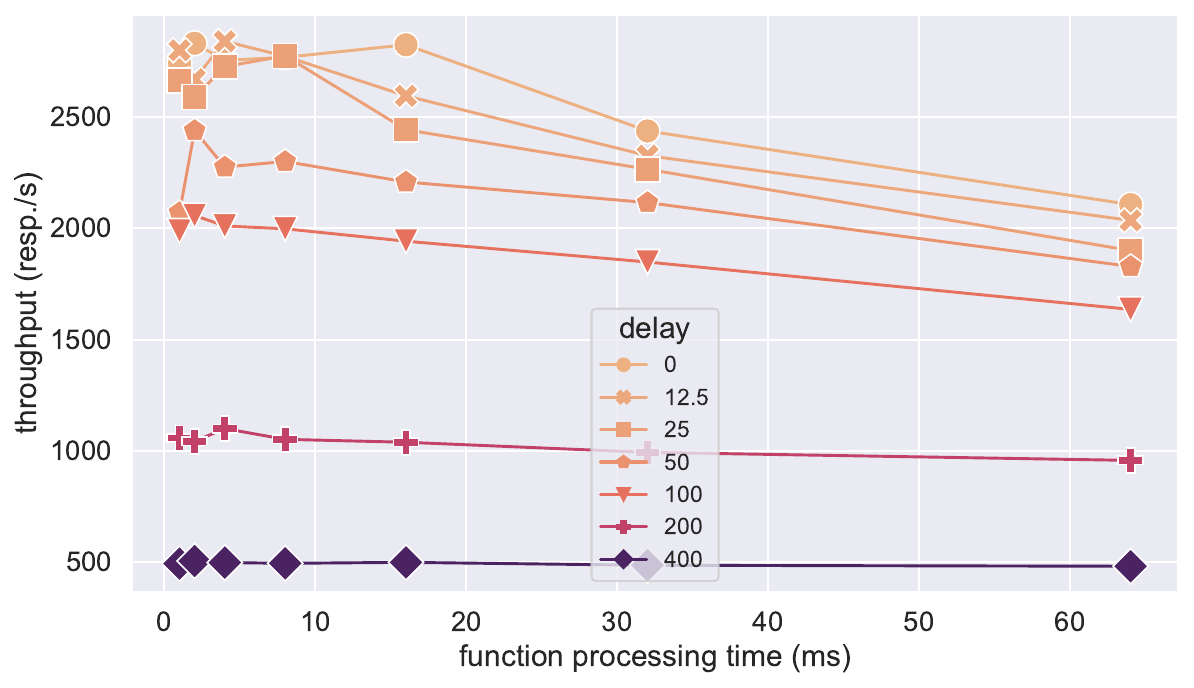}
\label{fig:scalability-sleep-fn_DS_T-500}} \caption{Throughput in number of
  successful requests/s versus processing time (ms)} \label{fig:scalability-sleep-fn}
\end{figure*}

\subsection{Tests}\label{sec:tests}

The specific user defined test is shown in Fig. \ref{fig:specific_diagram} and
can be described as follows.  In Step 6, the serverless platform Knative is
deployed on top of the Kubernetes cluster. From this point on, we start a loop
(Step 7) to iterate over all combinations of our two sets of test parameters;
namely thread number and processing time. For a set of $l$ users and $k$
different processing times, this results in a total of $k*l$ iterations per
delay-topology pair. For the number of concurrent test threads $l$ that are
simulating each user, we iterate over the set (1, 250, 500 and 1000),
while the set of processing times $k$ for the serverless function is (1, 2, 4,
8, 16, 32, 64 ms). Furthermore, the duration of all tests is set to 5 minutes.
The loop consists of installing Knative sleep function, running the test
with the load testing tool \emph{hey} and deleting the functions once the test
is done to avoid the influence of scaling operations of previous iterations on
the tests. The whole deployment and testing cycle is then repeated across
different a set of 7
different end-to-end delay values (0, 12.5, 25, 50, 100, 200 ,400 and 800
milliseconds), applied to both, the single site and multiple site configuration.

Every concurrent user of the \emph{hey} test creates a new request whenever it
receives a response for the preceding one until a time period defined by the
\emph{duration} parameter has passed. The \emph{load} parameter represents the
payload sent to the function, resulting in our simulated processing duration. We
define throughput as the number of successful requests per second, regardless of
the actual response times. The number of successful requests and the total
duration of the test are provided by \emph{hey}, which is then used to calculate
the throughput.

\section{Measurements and Performance }\label{sec:results}

We show the measurements of the achieved throughput versus the function processing times for
different delays. We compare the single site configuration versus multiple sites.  The presented
results are averaged over 5 to 10 different runs. Fig.  \ref{fig:scalability-sleep-fn} shows the
results where the x-axis represents the processing time of the function in ms, while the y-axis
represents the throughput in number of successful requests per second and each line plot shows the
emulated delay in ms introduced to the corresponding interfaces. 

Fig. \ref{fig:scalability-sleep-fn_CS_T-1} and ~\ref{fig:scalability-sleep-fn_DS_T-1} show the
results obtained when using a single virtual user on the tester machine. We can see there is great
similarity between the measurements for 25 ms of the single site and the 12.5 ms line
of the multi-site when looking at the load from just a single client; both of them start around 20
req/s at 0 ms processing time and then drop to 10 req/s at 65 ms processing time almost
identically. It can also be observed that the values decrease exponentially in both scenarios,
diminishing the significance of the inter-node latency with respect to absolute throughput numbers
the higher the processing time. For a single virtual user in the multi-site case with more than 50 ms of
delay, no relevant throughput could be measured. A possible explanation for this is the
intra-cluster communication in Knative that affects the multi-site configuration due to the delay
added in-between the cluster nodes, while this is not the case for a single site.

Fig. \ref{fig:scalability-sleep-fn_CS_T-50} and ~\ref{fig:scalability-sleep-fn_DS_T-50} show the
results for 50 concurrent users.  Also here the throughput decreases exponentially. For processing times of 0
to 60 ms, the multi-site throughput for 12.5 ms of inter-node delay ranges between the 25 and 50 ms
lines of the single site. For higher delays; 100 ms in the single site topology and 50 ms in the
multi-site, the throughput approaches an almost linear decrease with less incline the higher the
inter-node latency becomes. Overall, the multi-site scenario achieves only 30 - 40\% of the
throughput of the respective single site graph. This however changes slightly once the processing
duration increases, such that the multi-site 100 ms marker is positioned at roughly 125 req/s; 50\%
of the respective singe site marker at 250 req/s.

For 500 concurrent users in Figs.  \ref{fig:scalability-sleep-fn_CS_T-500} and
\ref{fig:scalability-sleep-fn_DS_T-500}, and up to 100 ms, the lines are much more similar in both
scenarios than observed in the previous measurements, highlighting a much lower impact of the
inter-node delay on the overall throughput with the distributed topology delivering around 90\% of
the performance of the single site.  For connections with higher delay, this however changes quickly
as can be observed by the significantly lower performance of the multi-site at 200 and 400 ms
of delay, where the performance again drops down to 40 - 60\% of the single site.  In the case of
multi-site, no results were produced for 800 ms which, we assume the cause for in the exceeding of
response timeouts.

In sum,  the measurements indicate that the throughput for the multi-site
environment is most of the time slightly above half of what it is for the single
site. This can be explained through the influence of requests that are served
directly by the headnode since no forwarding is necessary to handle them,
resulting in similar values as for a single site.

For applications without concurrency we can therefore observe a lower impact of
the topology on the throughput, meaning multi-site setups can meet roughly two thirds of the
performance of a single site once the processing duration exceeds 15 ms. For
50 concurrent users with low processing times, we can see roughly
half the single site performance for fast-responding applications with two thirds being
achieved when the processing time exceeds 30 ms. For 500 concurrent users, the effect of the processing time is much smaller than the effect
of the inter-node latency with almost comparable performance for delays lower
than 100 ms.

\section{Conclusions}

The presented an experimental study of the impact of inter-node delay impacts the application throughput. We
showed that this impact is significant and needs to be considered, even
if the processing time is significantly faster than the latency introduced. Regarding the
throughput, the impact of the delay introduced for 500 concurrent users however is significantly lower than for 50 concurrent users of less. We observe that there is need for 
a latency-aware scheduling process, in either Kubernetes or Knative tools in order to avoid sub-optimal
scheduling of containers across clusters with higher inter-node latency. An implementation
should possibly more beneficial in Kubernetes instead of Knative, since also non-serverless
applications likely are subject to the same consideration. Without a latency-aware scheduling plugin, Kubernetes and Knative may not be ready for multi-site distributed environments like in  the 
edge networking. Future work will consider various delay distribution in heterogeneous sites and the impact on application and system performance. 


\section*{Acknowledgment}
This work was partially supported by the project "Towards a functional continuum
operating system (ICOS)" funded by the European Commission under Project
code/Grant Number 101070177 through the HORIZON 2020 program.

\bibliographystyle{IEEEtran}
\bibliography{refs}

\end{document}